\documentclass[manuscript]{aastex}

\slugcomment{Submitted on 2013/Sep/10}

\shorttitle{A Discovery of a Candidate Companion to KOI-94}
\shortauthors{Takahashi et al.}

\begin{document}


\title{A Discovery of a Candidate Companion to a Transiting System KOI-94: A Direct Imaging Study for a Possibility of a False Positive}


\author{Yasuhiro H. Takahashi\altaffilmark{1,2,*}, Norio Narita\altaffilmark{2}, Teruyuki Hirano\altaffilmark{3}, Masayuki Kuzuhara\altaffilmark{3}, Motohide Tamura\altaffilmark{1,2}, Tomoyuki Kudo\altaffilmark{4}, Nobuhiko Kusakabe\altaffilmark{2}, Jun Hashimoto\altaffilmark{5}, Bun'ei Sato\altaffilmark{3}, Lyu Abe\altaffilmark{6}, Wolfgang Brandner\altaffilmark{7}, Timothy D. Brandt\altaffilmark{8}, Joseph C. Carson\altaffilmark{9},
Thayne Currie\altaffilmark{10}, Sebastian Egner\altaffilmark{4}, Markus Feldt\altaffilmark{7}, Miwa Goto\altaffilmark{11}, Carol A. Grady\altaffilmark{12,13,14}, Olivier Guyon\altaffilmark{4}, Yutaka Hayano\altaffilmark{4}, Masahiko Hayashi\altaffilmark{2}, Saeko S. Hayashi\altaffilmark{4}, Thomas Henning\altaffilmark{7}, Klaus W. Hodapp\altaffilmark{15}, Miki Ishii\altaffilmark{4}, Masanori Iye\altaffilmark{2}, Markus Janson\altaffilmark{8}, Ryo Kandori\altaffilmark{2}, Gillian R. Knapp\altaffilmark{8}, Jungmi Kwon\altaffilmark{2}, Taro Matsuo\altaffilmark{16}, Michael W. McElwain\altaffilmark{12}, Shoken Miyama\altaffilmark{17}, Jun-Ichi Morino\altaffilmark{2}, Amaya Moro-Martin\altaffilmark{8,18}, Tetsuo Nishimura\altaffilmark{4}, Tae-Soo Pyo\altaffilmark{4}, Eugene Serabyn\altaffilmark{19}, Takuya Suenaga\altaffilmark{2,20}, Hiroshi Suto\altaffilmark{2}, Ryuji Suzuki\altaffilmark{2}, Michihiro Takami\altaffilmark{21}, Naruhisa Takato\altaffilmark{4}, Hiroshi Terada\altaffilmark{4}, Christian Thalmann\altaffilmark{22}, Daigo Tomono\altaffilmark{4}, Edwin L. Turner\altaffilmark{8,23}, Makoto Watanabe\altaffilmark{24}, John Wisniewski\altaffilmark{5}, Toru Yamada\altaffilmark{25}, Hideki Takami\altaffilmark{2}, Tomonori Usuda\altaffilmark{2}}

\email{*; yasuhiro.takahashi@nao.ac.jp}


\altaffiltext{1}{Department of Astronomy, The University of Tokyo, 7-3-1, Hongo, Bunkyo-ku, Tokyo, 113-0033, Japan}
\altaffiltext{2}{National Astronomical Observatory of Japan, 2-21-1, Osawa, Mitaka, Tokyo, 181-8588, Japan}
\altaffiltext{3}{Department of Earth and Planetary Sciences, Tokyo Institute of Technology, Ookayama, Meguro-ku, Tokyo 152-8551, Japan}
\altaffiltext{4}{Subaru Telescope, National Astronomical Observatory of Japan, 650 North A`ohoku Place, Hilo, HI96720, USA}
\altaffiltext{5}{H.L. Dodge Department of Physics \& Astronomy, University of Oklahoma, 440 W Brooks St Norman, OK 73019, USA}
\altaffiltext{6}{Laboratoire Lagrange (UMR 7293), Universit\'e de Nice-Sophia Antipolis, CNRS, Observatoire de la C\^{o}te d'azur, 28 avenue Valrose, 06108 Nice Cedex 2, France}
\altaffiltext{7}{Max Planck Institute for Astronomy, K\"onigstuhl 17, 69117 Heidelberg, Germany}
\altaffiltext{8}{Department of Astrophysical Science, Princeton University, Peyton Hall, Ivy Lane, Princeton, NJ08544, USA}
\altaffiltext{9}{Department of Physics and Astronomy, College of Charleston, 58 Coming St., Charleston, SC 29424, USA}
\altaffiltext{10}{Department of Astronomy \& Astrophysics, University of Toronto, 50 George St., Toronto, Ontario, M5S 3H4, Canada}
\altaffiltext{11}{Universit\"ats-Sternwarte Mu\"nchen, Ludwig-Maximilians-Universit\"at, Scheinerstr. 1, 81679 Mu\"nchen, Germany}
\altaffiltext{12}{Exoplanets and Stellar Astrophysics Laboratory, Code 667, Goddard Space Flight Center, Greenbelt, MD 20771, USA}
\altaffiltext{13}{Eureka Scientific, 2452 Delmer, Suite 100, Oakland CA96002, USA}
\altaffiltext{14}{Goddard Center for Astrobiology}
\altaffiltext{15}{Institute for Astronomy, University of Hawaii, 640 N. A`ohoku Place, Hilo, HI 96720, USA}
\altaffiltext{16}{Department of Astronomy, Kyoto University, Kitashirakawa-Oiwake-cho, Sakyo-ku, Kyoto, Kyoto 606-8502, Japan}
\altaffiltext{17}{Hiroshima University, 1-3-2, Kagamiyama, Higashihiroshima, Hiroshima 739-8511, Japan}
\altaffiltext{18}{Department of Astrophysics, CAB-CSIC/INTA, 28850 Torrej\'on de Ardoz, Madrid, Spain}
\altaffiltext{19}{Jet Propulsion Laboratory, California Institute of Technology, Pasadena, CA, 171-113, USA}
\altaffiltext{20}{Department of Astronomical Science, The Graduate University for Advanced Studies, 2-21-1, Osawa, Mitaka, Tokyo, 181-8588, Japan}
\altaffiltext{21}{Institute of Astronomy and Astrophysics, Academia Sinica, P.O. Box 23-141, Taipei 10617, Taiwan}
\altaffiltext{22}{Astronomical Institute ``Anton Pannekoek'', University of Amsterdam, Postbus 94249, 1090 GE, Amsterdam, The Netherlands}
\altaffiltext{23}{Kavli Institute for Physics and Mathematics of the Universe, The University of Tokyo, 5-1-5, Kashiwanoha, Kashiwa, Chiba 277-8568, Japan}
\altaffiltext{24}{Department of Cosmosciences, Hokkaido University, Kita-ku, Sapporo, Hokkaido 060-0810, Japan}
\altaffiltext{25}{Astronomical Institute, Tohoku University, Aoba-ku, Sendai, Miyagi 980-8578, Japan}


\begin{abstract}
We report a discovery of a companion candidate around one of {\it Kepler} Objects of Interest (KOIs), KOI-94, and results of our quantitative investigation of the possibility that planetary candidates around KOI-94 are false positives. KOI-94 has a planetary system in which four planetary detections have been reported by {\it Kepler}, suggesting that this system is intriguing to study the dynamical evolutions of planets. However, while two of those detections (KOI-94.01 and 03) have been made robust by previous observations, the others (KOI-94.02 and 04) are marginal detections, for which future confirmations with various techniques are required. We have conducted high-contrast direct imaging observations with Subaru/HiCIAO in $H$ band and detected a faint object located at a separation of $\sim0.6''$ from KOI-94. The object has a contrast of $\sim 1\times 10^{-3}$ in $H$ band, and corresponds to an M type star on the assumption that the object is at the same distance of KOI-94. Based on our analysis, KOI-94.02 is likely to be a real planet because of its transit depth, while KOI-94.04 can be a false positive due to the companion candidate. The success in detecting the companion candidate suggests that high-contrast direct imaging observations are important keys to examine false positives of KOIs. On the other hand, our transit light curve reanalyses lead to a better period estimate of KOI-94.04 than that on the KOI catalogue and show that the planetary candidate has the same limb darkening parameter value as the other planetary candidates in the KOI-94 system, suggesting that KOI-94.04 is also a real planet in the system. 
\end{abstract}


\keywords{stars: imaging --- planets and satellites: individual (KOI-94) --- instrumentation: adaptive optics --- techniques: high angular resolution}



\section{Introduction}

One of the best ways to determine orbital parameters of extrasolar planets (exoplanets) is the transit method. Particularly the {\it Kepler} satellite, launched in 2009, has executed successful transit observations, resulting in discovery of more than 3,500 planet candidates\footnote{http://kepler.nasa.gov/}. It drastically increases the number of exoplanets we know and has found many multi-transiting planetary systems.

The principal problem related to the transit method is a possibility of a false positive. A false positive in transit surveys means misidentifying a signal caused by an object other than a planet orbiting the target star as its true planetary companion. In transit surveys, false positives are induced by some objects, for example eclipsing binaries, background transiting planetary systems or companions with transiting planets \citep{fre13} within photometric apertures of target stars. Such objects reduce their brightness periodically, and total flux containing both from the target stars and from the mimicking objects consequently decreases periodically. Since the presence of the false positive sources cannot be verified only with the transit method, the depressions in light curves cannot be translated directly into planets orbiting the targets, and follow-up observations are required to confirm that the depressions are really caused by planets.

Although most of {\it Kepler} planetary candidates are waiting to be confirmed, false positive rates reported by spectroscopic follow-up observations (\citealt{san12}; $\sim 35$\%) and other planet surveys (e.g. HAT-Net; \citealt{lat09}) are significantly higher than the rates for {\it Kepler} candidates theoretically expected by \citet{mor11} (less than 5\% for over half targets) and \citet{san13} ($11.3\pm1.1$\% for whole targets) based on Galactic models. The inconsistency has not been elucidated yet, and therefore, suggests that {\it Kepler} candidates are required to be confirmed with other manners, including the radial velocity method, TTVs \citep{ago05}, the centroid analysis \citep{bat10}, {\tt BLENDER} analysis \citep{tor04}, and the direct imaging observations.

Many follow-up observations with the direct imaging method have been executed so far, but most observations were too shallow to confirm candidates efficiently. For example, let us assume that there is a false positive source within the target star's photometric aperture with the magnitude of $\Delta m=10$ mag compared to the target. A full occultation of the source induces a depression with the depth of 100 ppm, which is a typical value caused by an Earth-like planet transiting a Sun-like star. A false positive source with a smaller contrast ($\Delta m<10$) can cause the 100 ppm depression by its partial occultation, while a source with a larger contrast ($\Delta m>10$) can induce a shallower depression. Thus, if a depression with the depth of 100 ppm is detected, direct imaging observations with a contrast of $\Delta m=10$ mag can fully and efficiently put a constraint on the possibility of a false positive. However, some studies employ direct imaging with a detection limit of $\Delta m<10$ mag for confirming a transit detection with the depth of $<100$ ppm (e.g. \citealt{bar13}). Such shallow observations cannot fully reject false positive sources. Moreover, if a companion candidate is found around the target by the direct imaging, we can evaluate gravitational influences upon the orbital migrations of the planetary system by the companion.

We focus on KOI-94, which was listed on the earliest {\it Kepler} Object of Interest (KOI) list \citep{bor11, bat12}. KOI-94 is a relatively faint --- $Kp=12.2$ \citep{bor11}, $H=11.0$ (2MASS; \citealt{skr06}) --- late F-type star with the age of $3.9^{+0.3}_{-0.2}$ Gyr \citep{hir12} or $3.16\pm0.39$ Gyr \citep{wei13}. This system has four planet candidates named as KOI-94.01, 02, 03 and 04, which are also known as KOI-94 d, c, e and b, respectively (properties listed on Table \ref{tbl-1}). \citet{hir12} discovered the ``planet-planet eclipse" phenomenon in the light curves, where the term means that a planet occults another planet transiting their host star at that time. Combining the event with measurements of the Rossiter-McLaughlin effect \citep{oht05} of KOI-94.01, they confirmed that KOI-94.01 and 03, at least, are real planets and showed that orbital axes of KOI-94.01 and 03 and the spin axis of the main star KOI-94 are well aligned. The fact suggests that the planets have not experienced the planet-planet scattering (e.g. \citealt{cha08,nag11}) or the Kozai migration (e.g. \citealt{koz62,wu07,fab07}). In contrast, a boundness of KOI-94.02 and 04 remains unclear; radial velocities observed by \citet{wei13} represented so low amplitudes that they could detect 04 only at a $2\sigma$ significance and 02 at the same level of non-detection, and their direct imaging observations had shallow depths ($\Delta K_S=3.4$ at 0\farcs5, 5.9 at $1\farcs0$) in spite of the depth of KOI-94.04 (131 ppm).
Hence, deeper direct imaging observations for KOI-94 are required to confirm detections of the candidates.

In this paper, we present results of high-contrast direct imaging observations for KOI-94 and elucidate the presence of a faint object around the star. Details of the direct imaging observations and discussion based on the observations are described in Section \ref{imaging-obs} and \ref{discussion-di}, respectively. We then show our results and discussion of reduced light curves of planetary candidates KOI-94.04 in Section \ref{discussion-lc}. Section \ref{summary} summarizes the paper.

\section{Deep direct imaging observations for KOI-94 and analyses\label{imaging-obs}}

In employing the direct imaging observations, we first estimate the distance of KOI-94 from the Earth. Given KOI-94's magnitudes of $g=12.551, i=12.057$ ({\it Kepler} Input Catalog) and metallicity of ${\rm [Fe/H]}=+0.0228\pm0.0020$ \citep{wei13} for KOI-94, we can infer its $r$-band absolute magnitude $M_r \sim 4.0$ according to Equation 1 in \citet{ive08}, who developed the Galactic model. A comparison between the $M_r$ and an observed magnitude of $r=12.186$ ({\it Kepler} Input Catalog) for KOI-94 enables us to estimate its distance to be $\sim 440$ pc. On the other hand, by applying an KOI-94's estimated mass of $1.25^{+0.03}_{-0.04}M_\odot$ \citep{hir12} to Yonsei-Yale isochrone model \citep{dem04}, we calculate $M_V\sim 3.9$, which can be compared with $V$-band magnitude of KOI-94 $V=12.6$ (NOMAD catalogue; \citealt{zac04}) to infer the distance of $\sim 550$ pc. The accurate distance does not matter for our following discussions, and we therefore adopt the average of both estimations, $\sim 500$ pc, in this study.

In order to check the presence of possible false positive sources, we observed the star as a part of the SEEDS project \citep{tam09}. The SEEDS has directly imaged stellar companions that are the important clue to the origin of close-in exoplanets \citep{nar10,nar12}, as well as substellar or planetary companions (e.g. \citealt{tha09,car13,kuz13}). We obtained $H$ band images using a high-contrast near-infrared camera HiCIAO \citep{hod08,suz10} with a 188 actuators adaptive optics system (AO188; \citealt{hay10}). The detector of HiCIAO consists of 2048$\times$2048 pixels with the plate scale of 9.5 mas/pixel, and has the field of view (FOV) of about $20'' \times 20'' $. The observations were performed on 2012 July 10\footnote{On the observation night, the eclipse of KOI-94.02 or 04 did not occur based on their transit information on the KOI catalogue.}, where we employed angular differential imaging (ADI; \citealt{mar06}) to remove the starlight and the stellar speckles. The sky was photometric, and the typical full width at half maximum (FWHM) of our observed PSFs was $\sim$15.4 pixels ($0\farcs146$) after AO corrections. We obtained 35 images, of which all frames were adequate for high-contrast science. Individual integration times were 15 seconds and 3 coadds per image were taken (i.e. each image has an integration time of 45 seconds). Thus our observations allow for the cumulative integration time of 26 minutes, where the field rotation through ADI observations was 13.9 degrees in total. We note that the observations were performed without an occulting mask and array saturation, which allow us to carefully calibrate the acquired data sets.

\begin{figure}
\plotone{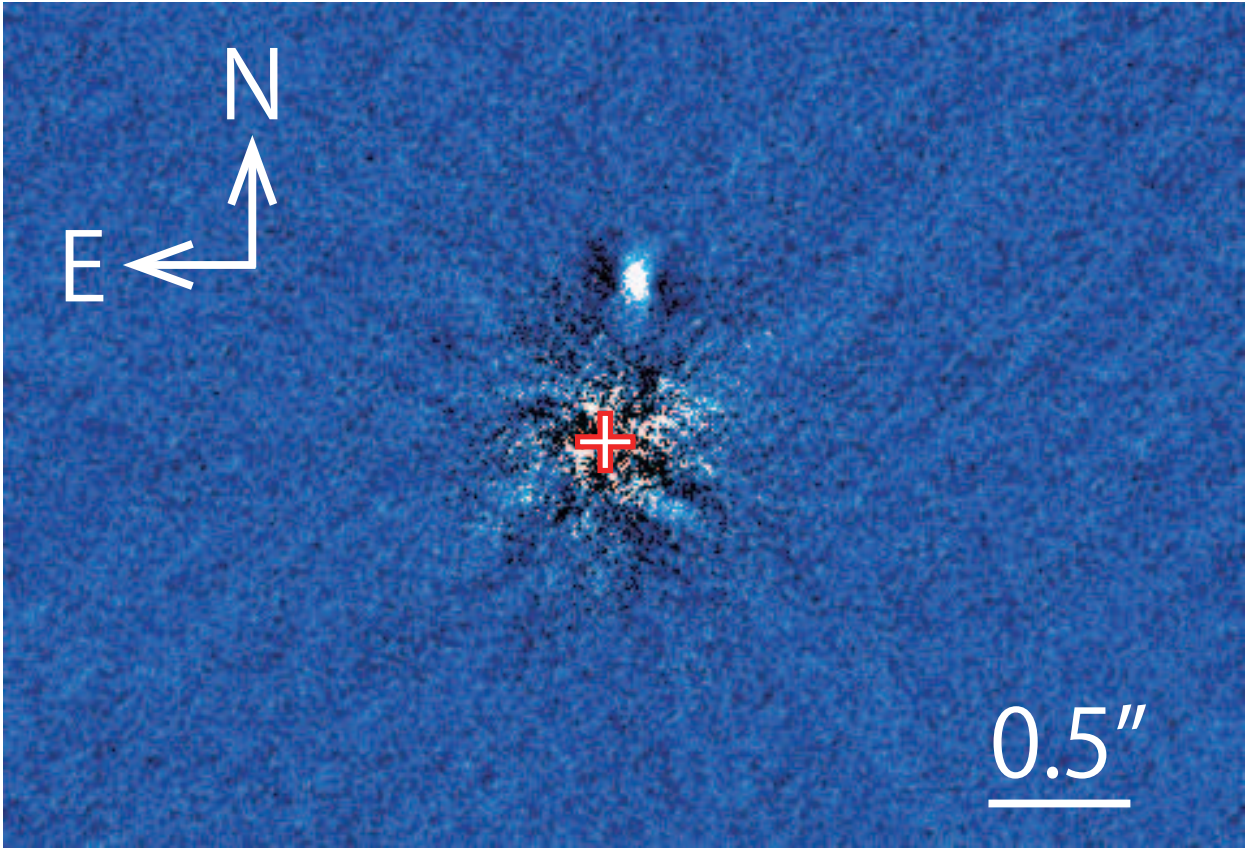}
\caption{An $H$ band image of KOI-94 reduced by the classical ADI analysis. The position of the central star is shown as a cross. A faint companion candidate appears at a separation of $\sim 0\farcs6$ from the star. Artificial dark tails extending east-west are induced by the ADI analysis.\label{fig-image}}
\end{figure}

\begin{figure}
\plotone{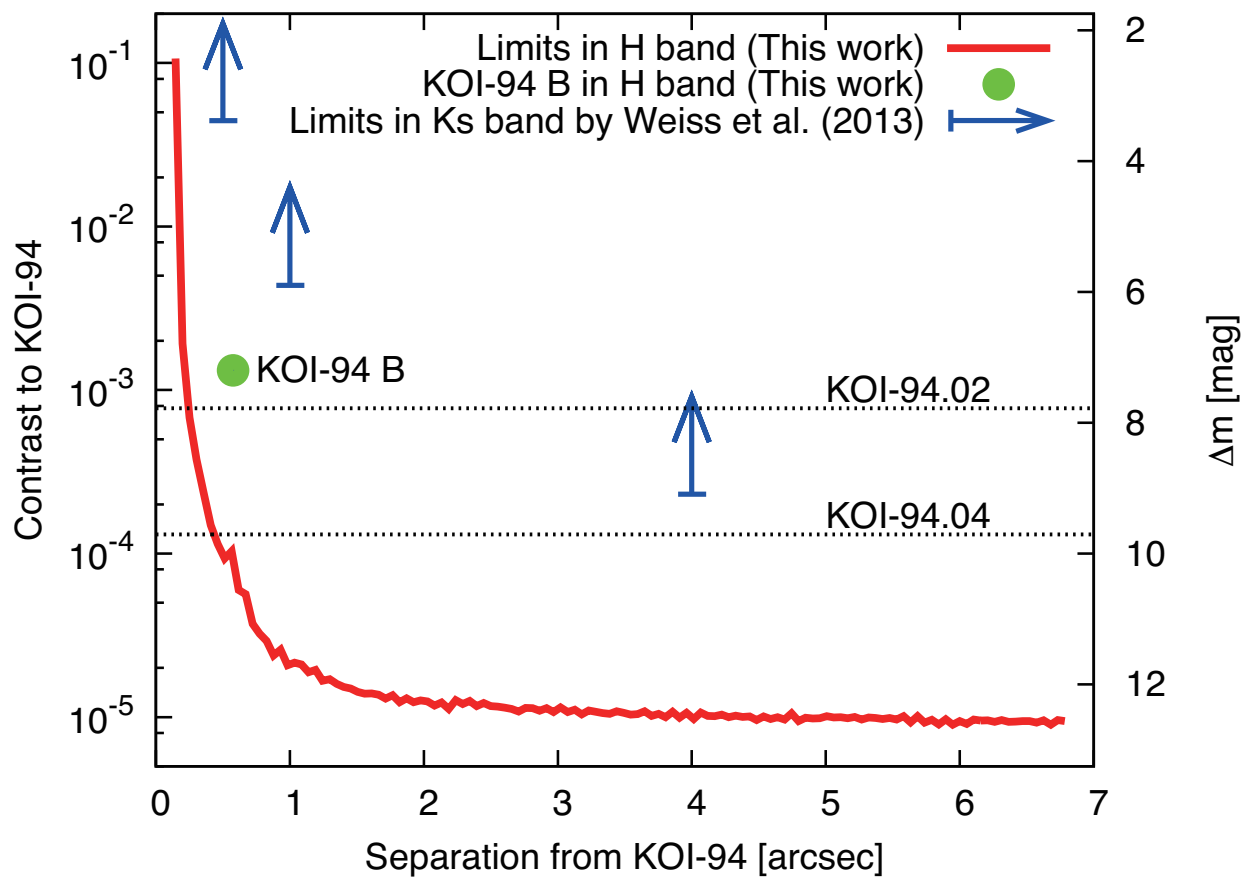}
\caption{A comparison between detection limits for our observations and the contrast of the companion candidate B, with the depths of KOI-94.02 and 04. The solid line shows a 5$\sigma$ contrast curve in $H$ band with Subaru/HiCIAO for KOI-94 overlaid with a position of KOI-94 B. The size of the circle is larger than its errors. The two dotted lines depict each depth of planetary candidates in $Kp$ band. The arrows represent limits by \citet{wei13} in $K_S$ band with MMT/ARIES for reference.\label{fig-contrast}}
\end{figure}

For the first attempt in data reductions, we plainly combined all frames after derotating them and subtracting halos from them, but spider noises and speckle noises prevented us from discussing the presence of faint objects around the target star. We then employed the locally optimized combination of images (LOCI) algorithm \citep{laf07} in order to remove the noises. As a result, we detect a faint object (hereafter KOI-94 B, or B for short) at a separation of $\sim 0\farcs6$ from KOI-94 (see Figure \ref{fig-image}), regardless of the fact that \citet{wei13} did not find the object with AO direct imaging observations using MMT/ARIES and the speckle imaging observations using DSSI camera on the WIYN 3.5m telescope. This is the first discovery of a false positive source candidate around KOI-94. The LOCI processing is powerful method for high-contrast data reductions, but it may not be efficient in the case of small field rotations as our observations for KOI-94 because of the serious self-subtraction problem\footnote{The self-subtraction is an inherent problem in the ADI analysis and means a diminution of a signal's flux together with various noises, induced by subtraction of the signal itself. In case that the signal is sufficiently bright and scarcely rotated, like KOI-94 B, LOCI tends to regard the signal as a noise and, consequently, LOCI attempts to decide coefficients too hard for the sake of removing the signal. As a result, the self-subtraction of LOCI can be larger than that of the classical ADI analysis.}, though some new techniques have been developed for avoiding the self-subtraction (e.g. \citealt{cur13}). Hence, we simply attempted to apply the classical ADI \citep{mar06} data processing to the images, after subtracting the halos of KOI-94 PSFs on each frame. Indeed, this additional attempt improved the signal-to-noise ratio (SNR) of KOI-94 B. A final image made by combining all frames reduced with the classical ADI analysis and its 5$\sigma$ contrast curve are shown in Figure \ref{fig-image} and \ref{fig-contrast}, respectively. We can easily find the faint star in Figure \ref{fig-image} at the position with a separation of $0\farcs6$ to the north (PA$=350^\circ$) from the central star. Figure \ref{fig-contrast} shows that our deep high contrast imaging (the red line) reaches $1.1\times 10^{-4}$ at $0\farcs5$ and $2.1\times 10^{-5}$ at $1\farcs0$, enabling us to detect B (shown by the green circle), in contrast to the shallow detection limits by \citet{wei13} (the blue arrows). The brightness of B was obtained from the aperture photometry relative to the central star ($H=11.0$), and compensated for the self-subtraction induced by the ADI analysis with estimates from measuring fluxes of artificial signals embedded in the science frames. Its photometric error is attributed mainly to the uncertainties of the estimates. Although the arrows are based on their brief descriptions of the detection limits in $K_S$ band, the difference of wavelengths does not affect the discussion as long as B has a moderate color of $H-K_S\lesssim 3$. The depths of KOI-94.02 and 04 in $Kp$ band are also represented in Figure \ref{fig-contrast} as the dotted lines, because we can directly compare them with B's contrast in $H$ band under condition that the transit depths are constant in various wavelengths. Properties of KOI-94 B are summarized in Table \ref{tbl-2} in detail.

Under an assumption that KOI-94 B is a single star with the same distance as that of KOI-94 ($\sim500$ pc), our measured magnitude, $H=18.2$, for KOI-94 can be converted into an absolute magnitude of $M_H=9.7$, which can be compared with 3-4 Gyr isochrones in NextGen model \citep{bar98} to infer a mass of $\sim 0.1 M_\odot$ for B. The measured projected separation of B ($\sim0\farcs6$) corresponds to $\sim300$ AU at the assumed distance.

\section{Interpretation of the results of the direct imaging observations\label{discussion-di}}

\subsection{Influences of the new companion upon the KOI-94 system\label{imaging-discussion}}

In this section, we examine influences from the companion candidate B found with the direct imaging.
First of all, we note that B does not correspond to any of the reported four planets orbiting KOI-94. This is because it is unrealistic that a planet with the separation of $\sim300$ AU (see Section 2) orbits the Sun-like star with the period of less than 54 days, the longest value in the KOI-94 system. Since the most important fact is that the candidate B is within the {\it Kepler} photometric aperture (typically a few arcsecs), we discuss influences for all hypotheses upon the light curves of KOI-94 in this section. Among the depressions of KOI-94, because we know that KOI-94.01 and 03 are real planets transiting KOI-94 \citep{hir12,wei13}, only KOI-94.02 and 04 are to be discussed.

The following cases can cause depressions in the light curves of KOI-94.02 or 04; (1) a real planet transiting the central star KOI-94, (2) a real planet transiting the companion B, and (3) B being an eclipsing binary. In case of (1), B cannot affect estimates of stellar and planetary parameters from the light curves seriously, because B is much fainter than the central star. In cases of (2) and (3), whether B is bound or not to the central star KOI-94 does not become a serious issue. If (2) or (3) is true, planet candidates of KOI-94.02 or 04 may not be a real planet orbiting KOI-94. These hypotheses about KOI-94.02 are examined and concluded in the following paragraphs based on the direct imaging observations. The hypotheses (2) and (3) about KOI-94.04 are precisely discussed in \S 3.2 in addition to this subsection, while the hypothesis (1) about it is argued in \S 4.2 grounded on our light curve reanalyses (\S 4.1).

Considering an extreme case that the companion candidate is completely occulted, the brightness of B can be translated into upper limit for the depth of depression induced by B. Then if the upper limit is shallower than the transit depth of KOI-94.02 or 04, it is impossible that KOI-94.02 or 04 is orbiting the companion candidate, i.e., it is a real planet transiting the central star. In order to estimate B's {\it Kepler} magnitude from our $H$ band magnitude and directly compare it with the depths of the depressions of the planetary candidates in $Kp$ band on the KOI catalogue, we adopt the following analyses.

First, if B is physically bound to KOI-94 and a single star, B's mass should be $\sim0.1M_\odot$ (see Section 2) and we can then calculate its apparent brightness in $Kp$ band to be $Kp=24.0$ using estimated absolute magnitudes of $M_g=16.9$ and $M_r=15.4$ \citep{bar98} from the observed magnitude in $H$ band and the equation\footnote{http://keplergo.arc.nasa.gov/CalibrationZeropoint.shtml}
\begin{equation}
Kp=\left\{
\begin{array}{ll}
0.2 g+0.8 r & {\rm for}\ g-r\leq 0.8\\
0.1 g+0.9 r & {\rm for}\ g-r>0.8
\end{array}\right.,
\end{equation}
where $g$ and $r$ are apparent magnitudes for an apparent $Kp$ magnitude. Consequently we obtain $\Delta Kp=11.8$, corresponding to a contrast of $1.9\times 10^{-5}$ in $Kp$ band. The contrast, which means an upper limit, is too small to explain the depths of KOI-94.02 ($7.6\times 10^{-4}$) or 04 ($1.3\times 10^{-4}$). Therefore, in this case, hypothesis (2) is ruled out.

Simultaneously, the hypothesis (3) that B is an eclipsing binary bound to KOI-94 can be discussed. Because each component of the binary would be fainter than the single M star, the binary would consist of later type stars than the above estimate. The later stars can produce shallower depressions in the light curves, and thus the hypothesis (3) is also excluded.

Next we investigate the situation that B is not bound to KOI-94. In this case, the B's color cannot be constrained from our measured $H$-band magnitude and the theoretical evolutionary trucks, since it is sufficiently possible that B has the distance and age different from those of KOI-94. Thus we are not able to directly estimate the $Kp$ magnitude. We employ statistical discussions to estimate it based on \citet{howe12}, who compared {\it Kepler} Input Catalogue with 2MASS catalogue and derived empirical relationships between infrared magnitudes and $Kp$ magnitudes of stars in the {\it Kepler} field. Since they did not indicate $Kp$-$H$ relationship, we substitute our $H$ magnitude into $J$ and $K_S$ magnitudes in their equations in order to obtain a rough estimate of B's $Kp$ magnitude. Then we acquire $Kp=19.9$ and $Kp=21.5$ with $Kp$-$J$ and $Kp$-$K_S$ relationships, respectively. Since the magnitude of KOI-94 central star is $Kp=12.2$, the contrasts are $\Delta Kp=7.7$ or 9.3. These contrasts correspond to flux ratios of $8.3\times 10^{-4}$ and $1.9\times 10^{-4}$, respectively. Comparing them with the depth of KOI-94.02 ($7.8\times 10^{-4}$), we suggest that it is difficult to regard KOI-94.02 as a planet transiting B or a false positive induced by B; in other words, an extreme case that B is completely occulted can explain the depth of KOI-94.02. On the other hand, the possibility that KOI-94.04 ($1.3\times10^{-4}$) is a false positive cannot be excluded by comparing B's magnitude and its depth. Hence we conclude that KOI-94.02 is a real planet orbiting KOI-94, assuming the typical color of stars in the field, and discuss KOI-94.04's nature in the following sections.

\subsection{Is KOI-94.04 a False Positive?\label{discussion-fp}}

In this subsection, we assume KOI-94.04 as a false positive.

First we examine the hypothesis (2) that the planetary candidate orbits B. Because the assumption that the companion B is physically bound to the central star is rejected in \S \ref{imaging-discussion}, we here assume the companion candidate to be a background star.

Comparing the depth of planetary candidate KOI-94.04 of $1.3\times 10^{-4}$ with the contrast of B of $(1.9-8.3)\times 10^{-4}$ in $Kp$ band (see \S 3.1), a ratio of their radii is $R_{\rm KOI-94.04}/R_{\rm B}=0.4-0.8$, if any light from KOI-94.04 is neglected. The estimation of $R_{\rm{KOI-94.04}}$/R$_{\rm{B}}$ can constrain the radius of KOI-94.04. If B has a radius larger than $0.5R_\odot$, the ratio leads to $R_{\rm KOI-94.04} > 2.0 R_{\rm J}$; an exoplanet with a radius larger than 2.1 $R_{\rm{J}}$ has not been discovered so far (\citealt{wei13}, see also exoplanets.org). Considering the estimate of $R_{\rm KOI-94.04}/R_{\rm B}$ and the observational knowledge cumulative for exoplanet properties, we can suggest that a star with a radius of $>0.5R_\odot$ can hardly explain the depressions for KOI-94.04. Note, however, an object with $R_{\rm B}< 0.5R_\odot$ can account for our estimated $R_{\rm KOI-94.04}/R_{\rm B}$. Obtaining a spectral type of B and determining its radius by future observations are important for investigating the possibility of the hypothesis (2).

Secondly we investigate the hypothesis (3) that the companion candidate B is a binary. B's contrast in $Kp$ band and the depth of KOI-94.04 are the same as those in the above situation. If the amount of B's fluxes decreases by 20-70\% via the eclipse in the system, the depth can be explained, and thus the possibility of B being a binary is not rejected. Moreover, \citet{fre13} and \citet{san13} calculated the probability of false positive with a given {\it Kepler} transit depth, leading to the estimated false positive rate of $8.9\pm2.0$\% for KOI-94.04 in combination of their calculations. Consequently, the possibility of a false positive remains for KOI-94.04. If the secondary eclipses of KOI-94.04 is detected, it becomes strong evidence supporting KOI-94.04 as an eclipsing binary. However, we did not find a significant secondary eclipse in the following light curve analyses. Thus the possibility that KOI-94.04 is a planet (i.e. the hypothesis (1)) remains, which is discussed in the next section.

\section{Constraints by KOI-94 light curves\label{discussion-lc}}

Because we cannot fully exclude the possibility that KOI-94.04 is a false positive even with the new direct imaging observations, we investigate the possibility of KOI-94.04 being a real planet by revisiting the KOI-94.04 transit light curves. \S 4.1 shows our reductions and results, and we discuss the possibility in \S 4.2 based on \S 4.1.

\subsection{Reanalysis and results of light curves}

We here employ {\it Kepler}'s public data sets \citep{bor11,bat12} of Quarter 1 through Quarter 13. In order to detect KOI-94.04 in the light curves, we first remove trends on the curves by fitting polynomial functions with masking transit depressions. Second we fold the curves by KOI-94.04's period on the KOI catalogue (3.743245 days in \citealt{bor11}). Third we bin the folded light curves into thousand data points. Errors of the binned points are based on the scatter of the data points in each bin. As a result, we see a depression of KOI-94.04.

\begin{figure}
\plotone{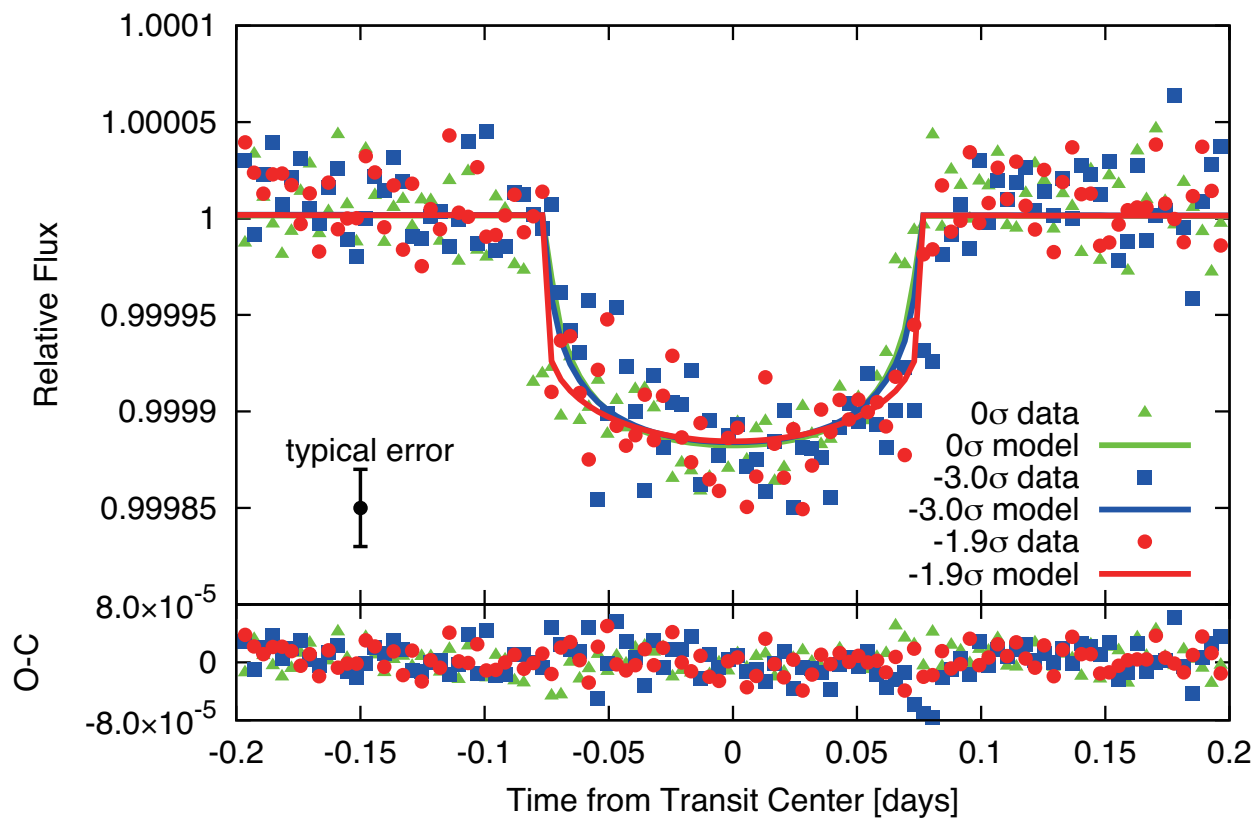}
\caption{Three KOI-94.04 light curves phase-folded by three periods as a function of time from the transit centers. The dots and lines in the upper panel represent observational and modeled relative flux data. Typical error size for each dot is shown at the lower left. The colors reflect differences of KOI-94.04's period; $0\sigma$ (i.e. not shifted), $-3.0\sigma$ and $-1.9\sigma$ (the lowest $\chi^2$) shifted from the KOI-catalogued period are green, blue and red, respectively. Lower panel shows residuals.\label{fig-lightcurve}}
\end{figure}

\begin{figure}
\plotone{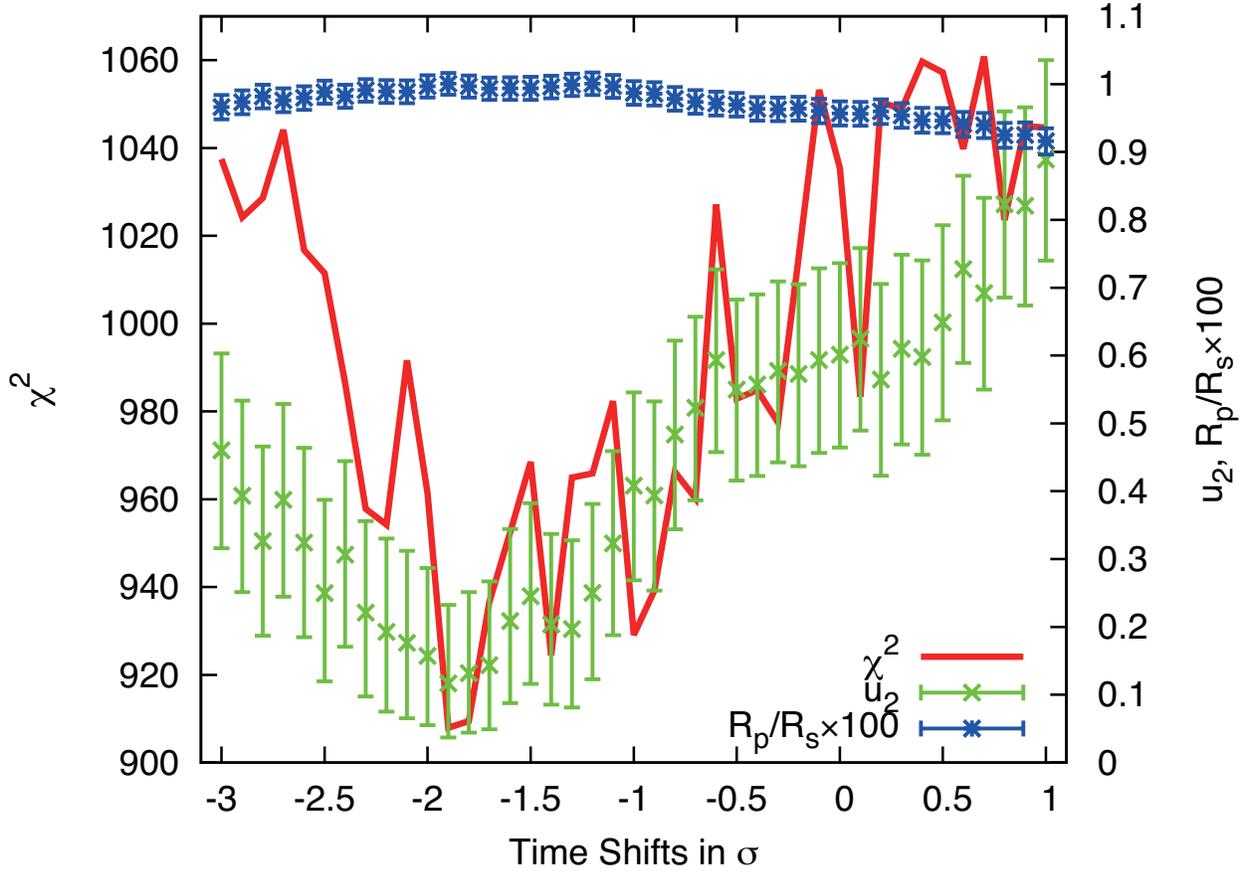}
\caption{Variations of parameters for KOI-94.04 as a function of its period. The red line indicates variation of the $\chi^2$ values. The green and blue dots with error bars represent the limb-darkening $u_2$ parameter and the radial ratio between the planet (KOI-94.04) and the star (KOI-94) multiplied by 100. The horizontal axis shows period shifts from the KOI-catalogued period in a unit of its error. \label{fig-chi2}}
\end{figure}

The shape of phase-folded transit light curve looks asymmetric, suggesting that the fiducial period may be incorrect. Therefore, we systematically changed its period from the catalogued value with shifts in unit of its error (0.000031 days) to find a symmetric transit light curve, because a folded transit light curve with an exact period would give a symmetric curve. Figure \ref{fig-lightcurve} shows three light curves with various periods and their best fit model curves. The $-1.9\sigma$ light curve (i.e. $P=3.743245-1.9\times 0.000031=3.743186$ days) gives the most symmetric light curve. We adopt transit model by \citet{oht09}, which uses the quadratic limb-darkening law, defined as
\begin{equation}
I(\mu)=1-u_1(1-\mu)-u_2(1-\mu)^2,
\end{equation}
where $I$ is the intensity and $\mu$ is the cosine of the angle between the line of sight and the line from the stellar center to the position of the stellar surface.
In order to quantify their symmetricities, we measured a $\chi^2$ parameter obtained by fitting a transit model function to the curves. Here, the $\chi^2$ parameter is expressed as
\begin{equation}
\chi^2=\sum_i \frac{(F_{{\rm model}, i}-F_{{\rm obs}, i})^2}{\sigma_i^2},
\end{equation}
where $F_{{\rm model}, i}$ and $F_{{\rm obs}, i}$ are the $i$-th modeled and observed relative flux data, and $\sigma_i$ is its error.  The curves other than $-1.9\sigma$ shifted data sets resulted in the higher $\chi^2$ parameters. In addition to $\chi^2$, limb-darkening parameter $u_2$ and their radial ratio between the planet candidate KOI-94.04 and its host star KOI-94, $R_p/R_s$, are depicted in Figure \ref{fig-chi2} as a function of time shift. In the reduction, we fixed the other limb-darkening parameter $u_1$ at 0.40 (Masuda et al. in prep.) and see the variation of the best-fit $u_2$ value. The light curve shifted by $-1.9\sigma$ (the red dots and line in Figure \ref{fig-lightcurve}) indicates the smallest $\chi^2$ value in our grid survey (cf. Figure \ref{fig-chi2}), which means 3.743186 days gives the best for its period. The corresponding $u_2$ and $R_p/R_s$ are $0.12^{+0.12}_{-0.08}$ and $0.01001\pm0.00016$, respectively. This estimate of $u_2$ is especially used for the confirmation of planetary candidate KOI94.04 in the following section.

\subsection{Is KOI-94.04 a Planet?}

In contrast to \S \ref{discussion-fp}, we investigate the possibility of KOI-94.04 as a real planet orbiting KOI-94 --- the hypothesis (1). \citet{wei13} conducted spectroscopic observations with Keck/HIRES, putting limits on the masses of KOI-94.02 and 04. They showed KOI-94.04's period of 3.74 days, scaled semi-major axis of $a/R_*=7.25\pm 0.59$, eccentricity of $0.25\pm 0.17$, radius of $1.71\pm 0.16R_\oplus$, and mass of $10.5\pm 4.6M_\oplus$. Surprisingly, the radius and the mass lead to an extraordinarily high density of $10.1\pm 5.5 {\rm \ g\ cm^{-3}}$. Also, they determined that KOI-94.02 has a period of 10.42 days and eccentricity of $0.43\pm0.23$. Meanwhile, according to their orbital stability analysis, 80\% of the random simulations with eccentric orbits reached close encounters between KOI-94.04 and KOI-94.02. Namely eccentric orbits typically become dynamically unstable in the KOI-94 system in their analysis, and they therefore suggested that the planets should have circular orbits to keep the system dynamically stable. Nevertheless their radial velocity data support the eccentric orbits for KOI-94.02 and 04. This inconsistency allows us to consider the mass or eccentricity estimate for KOI94.02 or 04 possibly inaccurate. Furthermore, though \citet{wei13} also showed TTVs in the KOI-94 system, the TTV of KOI-94.04 was not discussed in their TTV analysis. Hence, the possibility that KOI94.04 is a false positive cannot be ruled out by the discussions of \citet{wei13}.

Our limb-darkening analyses also enable us to discuss whether the planetary candidates orbiting the same host star or not by comparing them. Our analyses show that the limb darkening parameter $u_2$ for the $-1.9\sigma$ shifted data (i.e. the lowest $\chi^2$) is $u_2=0.12^{+0.12}_{-0.08}$, under the condition that the $u_1$ parameter is fixed at 0.40. Our values of $u_1$ and $u_2$ are consistent with those of KOI-94.01, 02 and 03 reported by Masuda et al. (in prep; $u_1\sim0.40$ and $u_2\sim0.14$), who analyzed $u_1$ and $u_2$ simultaneously as free parameters. The consistency supports that KOI-94.04 is orbiting the same host star. In contrast, the $u_2$ parameter of the data sets phase-folded with the catalogued period (i.e. $0\sigma$) is $u_2=0.60^{+0.13}_{-0.14}$, which is inconsistent with those of the other planets. The discrepancy implies KOI-94.04 does not transit KOI-94, but its $\chi^2$ parameter is worse than that of $-1.9\sigma$ data sets. The $\chi^2$ parameter suggests that true KOI-94.04's period may not be the value on the catalogue but slightly shifted (our result $P=3.743186$ days is compatible with that of \citealt{wei13}, $P=3.743208\pm0.000015$ days), and KOI-94.04 is a possible planet orbiting KOI-94 based on the limb-darkening parameters. Although the results are still imperfect for the confirmation, the discussions in this section contribute to improving the limits on the possibility of a false positive of KOI-94.04.

Apart from the problem of false positives for KOI-94, B may induce gravitational interaction with the planets in the system under the assumption that B is physically bound to KOI-94. The interaction can make the planets migrate and intersect their orbits, leading the system unstable.

\section{Summary\label{summary}}

We have focused on the multiple planetary system KOI-94 and conducted high-contrast direct imaging observations with Subaru/HiCIAO in order to examine a possibility of false positives. As a result of our classical ADI analysis, we have discovered a faint object at a separation of 0.6 arcsecs with a contrast of $\Delta H\sim 1 \times 10^{-3}$ to the north. Our estimates of its magnitude in $Kp$ revealed that the insubstantial binary in the background can explain the depths of KOI-94.04 as a false positive. We have also excluded the possibility of a false positive of KOI-94.02 because the depths of KOI-94.02 is almost equal to or larger than the contrast of the faint object, assuming the companion candidate being a real companion or a background star with the typical color. On the other hand, our transit analyses show the limb darkening parameter of KOI-94.04 is consistent with those of the other planetary candidates in the KOI-94 system, suggesting that KOI-94.04 might orbit the same host star. Although we cannot conclude that KOI-94.04 is a planet from our results, we have demonstrated that the combination of the direct imaging and analysis of transit light curve can constrain the possibility of a false positive. 

Our results also suggest that it is not enough to conclude the possibility of false positives by shallow or low-contrast direct imaging observations, which have been often conducted so far. Considering the fact that \citet{wei13} failed to find the companion candidate B due to their detection limits, our results suggest to require deep observations with $\Delta m\gtrsim10$ mag at a few arcsecs for the confirmation of a planet with its transit depth of $\lesssim100$ ppm rather than conventional and shallow observations for excluding false positives. Furthermore, the discovery of the companion candidate B, if confirmed to be bound, gives an important clue to dynamical evolution of the planets in the KOI-94 system.

\acknowledgments

This paper is based on data collected at the Subaru telescope, operated by National Astronomical Observatory of Japan. We thank the special support for HiCIAO and AO188 observations by the staffs. We are also grateful to NASA's {\it Kepler} Mission, which obtained photometric data of KOI-94. The data analysis was operated on common use data analysis at the Astronomy Data Center, ADC, of the National Astronomical Observatory of Japan. The works by Y.H.T. and T.H. are supported by Japan Society for Promotion of Science (JSPS) Fellowship for Research (DC1: 23-271, No.~25-3183). N.N. acknowledges supports by NAOJ Fellowship, NINS Program for Cross-Disciplinary Study, and Grant-in-Aid for Scientific Research (A) (No.~25247026) from the Ministry of Education, Culture, Sports, Science and Technology (MEXT) of Japan. This work is partly supported by the JSPS fund (No.~22000005). The work by J.C.C. is supported by the U.S. National Science Foundation under Award No.~1009203.




\clearpage



\clearpage

\begin{table}
\begin{center}
\caption{Properties of planetary candidates in KOI-94\label{tbl-1}}
\begin{tabular}{rcccc}
\tableline\tableline
Candidate & Period (days) & Depth (ppm) & $M_{\rm pl}$ ($M_\oplus$) & $R_{\rm pl}$ ($R_\oplus$) \\
\tableline
KOI-94.01 (d) & $22.343000\pm0.000011$& 5747 & $106\pm11$ & $11.27\pm1.06$\\
02 (c) & $10.423707\pm0.000026$ & 775 & $15.6^{+5.7}_{-15.6}$ & $4.32\pm0.41$\\
03 (e) & $54.31993\pm0.00012$ & 1932 & $35^{+18}_{-28}$ & $6.56\pm0.62$\\
04 (b) & $3.743245\pm0.000031$ & 131 & $10.5\pm4.6$ & $1.71\pm0.16$\\
\tableline
\end{tabular}

The periods and depths are in \citet{bor11}, and the others are in \citet{wei13}.
\end{center}
\end{table}

\clearpage

\begin{table}
\begin{center}
\caption{Our result of KOI-94 B\label{tbl-2}}
\begin{tabular}{cccc}
\tableline\tableline
Separation ($''$) & Position Angle (deg) & Contrast ($\times 10^{-4}$) & $\Delta H$ (mag)\\
\tableline
$0.5733\pm 0.0043$ & $349.90\pm 0.34$ & $13.2\pm 2.3$ & $7.2\pm 0.2$ \\
\tableline
\end{tabular}

Contrast is relative to the central star, KOI-94 A. $\Delta H$ is the $H$ band flux ratio to the central star.
\end{center}
\end{table}

\clearpage

\clearpage

\clearpage

\clearpage

\end{document}